\documentclass[english,twocolumn,10pt, A4paper]{IEEEtran}
\usepackage[T1]{fontenc}
\usepackage[latin9]{inputenc}
\usepackage{bm}
\usepackage{amsmath,amssymb,bm,color}
\usepackage{graphicx}
\usepackage{esint}

\makeatletter
\usepackage{bm}
\usepackage{subfigure}
\usepackage{cite}
\usepackage[english]{babel}
\usepackage[T1]{fontenc}
\usepackage{algorithm}
\usepackage{algorithmic}
\makeatletter
\def\blfootnote{\xdef\@thefnmark{}\@footnotetext}
\makeatother

\makeatother

\usepackage{babel}
\begin{document}

\title{Delay-Optimal Computation Task Scheduling for Mobile-Edge Computing Systems}
\author{\IEEEauthorblockN{Juan Liu{$^\dagger$}, Yuyi Mao{$^\dagger$}, Jun Zhang{$^\dagger$}, and K. B. Letaief{$^{\dagger\ast}$}, \emph{Fellow, IEEE}}\\
\IEEEauthorblockA{$^\dagger$Dept. of ECE, The Hong Kong University of Science and Technology, Clear Water Bay, Hong Kong\\
$^{\ast}$Hamad bin Khalifa University, Doha, Qatar\\
Email: eeliujuan@gmail.com, \{ymaoac, eejzhang, eekhaled\}@ust.hk}
}
\maketitle
\begin{abstract}
Mobile-edge computing (MEC) emerges as a promising paradigm to improve the quality of computation experience for mobile devices. Nevertheless, the design of computation task scheduling policies for MEC systems inevitably encounters  a challenging two-timescale stochastic optimization problem. Specifically,  in the larger timescale, whether to execute a task locally at the mobile device or to offload a task to the MEC server for cloud computing should be decided, while in the smaller timescale, the transmission policy for the task input data should adapt to the channel side information. In this paper, we adopt a Markov decision process approach to handle this problem, where the computation tasks are scheduled based on the queueing state of the task buffer, the execution state of the local processing unit, as well as the state of the transmission unit.  By analyzing the average delay of each task and the average power consumption at the mobile device, we formulate a power-constrained delay minimization problem, and propose an efficient one-dimensional search algorithm to find the optimal task scheduling policy. Simulation results are provided  to demonstrate the capability of the proposed optimal stochastic task scheduling policy in achieving a shorter average execution delay compared to the baseline policies.
\end{abstract}


\section{Introduction}
With the proliferation of smart mobile devices, computationally intensive applications, such as online gaming, video conferencing and 3D modeling, are becoming prevalent. However, the mobile devices normally possess limited resources, e.g., limited  battery energy and  computation capability of  local CPUs, and thus may suffer from unsatisfactory computation experience. Mobile-edge computing (MEC) emerges as a promising remedy. By offloading the computation tasks to the physically proximal MEC servers, the quality of computation experience, e.g., the device energy consumption and the execution delay, could be greatly improved \cite{Kumar12,ETSI}.

Computation offloading policies play critical roles in MEC,  and  determine  the efficiency and achievable computation performance \cite{Barbarossa1411}. Specifically, as computation offloading requires wireless data transmission, optimal computation offloading policies should take the time-varying wireless channel into consideration. In \cite{DHuang1206}, a stochastic control algorithm adapted to the wireless channel condition was proposed to decide the offloaded software components. A game-theoretic computation offloading approach for multi-user MEC systems was proposed in \cite{XChen1510}, and this study was extended to multi-cell settings in \cite{Sardellitti1506}. Besides, the energy-delay tradeoff in cloud computing systems with heterogeneous types of computation tasks and multi-core  mobile devices was investigated using Lyapunov optimization techniques in \cite{JKwak1512} and \cite{ZJiang1512}, respectively. In addition, a dynamic computation offloading policy for MEC systems with mobile devices powered by renewable energy was developed in \cite{YMao1612}.

For most mobile applications, the execution time is in the range of tens of milliseconds, which is much longer than the time duration of a channel block, whose typical value is a few milliseconds. In other words, the execution process may experience multiple channel blocks, which makes the computation offloading policy design a highly challenging two-timescale stochastic optimization problem. In particular, in a larger timescale, whether to offload a task to the MEC server or not needs to be decided, while in a smaller timescale, the transmission policy for offloading the input data of an application should adapt to the instantaneous wireless channel condition. To handle this issue, an initial investigation for two-timescale computation offloading policy design was conducted in \cite{WZhang1309}, which, however, only considered to minimize the energy consumption of executing a single computation task and the queueing delay incurred by multiple tasks was ignored. Moreover, with MEC, the potential of executing multiple tasks concurrently should be fully exploited in order to utilize the local and cloud computation resources efficiently and improve the quality of computation experience to the greatest extent.

In this paper, we will investigate an MEC system that allows parallel computation task execution at the mobile device and at the MEC server. The execution and computation offloading processes of the computation tasks running at the mobile device may be   across  multiple channel blocks, and the generated but not yet processed tasks are waiting in a task buffer. The average delay of each task and the average power consumption at the mobile device under a given computation task scheduling policy are first analyzed using Markov chain theory. We then formulate the power-constrained delay minimization problem. An efficient one-dimensional search algorithm is developed to find the optimal stochastic computation offloading policy. Simulation results show that the proposed stochastic computation task scheduling policy achieves substantial reduction in the execution delay compared to the baseline schemes.

The rest of this paper is organized as follows. We introduce the system model in Section II. The average execution delay and the power consumption of the mobile device under a given stochastic computation task scheduling policy are analyzed in Section III. In Section IV, a power-constrained delay minimization problem is formulated and the associated optimal task scheduling policy is obtained. Simulation results are shown in Section V and conclusions are drawn in Section VI.

\section{System Model\label{sec:System-Model}}

\begin{figure}[tp]
\includegraphics[width=0.48\textwidth]{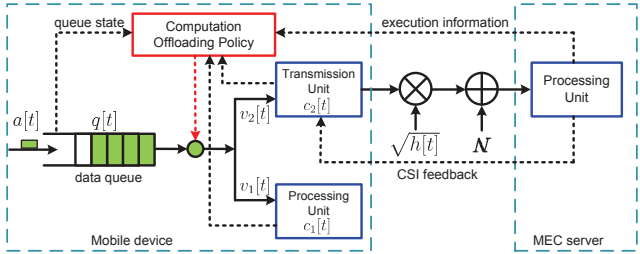}
\protect\caption{An MEC system with a mobile device and an MEC server. \label{fig:System-model}}
\end{figure}

We consider a mobile-edge computing (MEC) system as shown in Fig. \ref{fig:System-model}, where a mobile device is running computation-intensive and delay-sensitive applications with the aid of an MEC server. This MEC server could be a small data center installed at the wireless access point. By constructing a virtual machine associated with the mobile device, the MEC server can execute the computation tasks on behalf of the mobile device \cite{WZhang1309}. The CPU and the transmission unit (TU) at the mobile device are of particular interests, which can execute the computation tasks locally and transmit the input data of the computation tasks to the MEC server for cloud computing, respectively. Besides, due to the limited battery size and in order to prolong the device lifetime,  we assume that the average power consumption at the mobile device is constrained by $\bar{P}_{\max}$.

\subsection{Task Queueing Model}
We assume that time is divided into equal-length time slots and the time slot length is denoted as $\Delta$. At the beginning of each time slot, with probability $\alpha$, a new task is generated. The computation tasks can either be executed at the mobile device by the local CPU or be offloaded to the MEC server for cloud computing. The arrived but not yet executed tasks will be queued in a task buffer with a sufficiently large capacity $Q$\footnote{With this assumption, buffer overflows will not happen and all the arrived tasks will be executed either by the local CPU or by the MEC server.}. Denote $v_{L}\left[t\right],v_{C}\left[t\right]\in\{0,1\}$ as the computation task scheduling decision indicators for the $t$th time slot, i.e., if a task is decided to be sent to the local CPU (MEC server) in the $t$th time slot, $v_{L}\left[t\right]=1$ ($v_{C}\left[t\right]=1$); otherwise $v_{L}\left[t\right]=0$ ($v_{C}\left[t\right]=0$). Thus, there  are  four  possible computation task scheduling decisions, i.e., $\mathcal{V}=\{\left(v_{L}\left[t\right],v_{C}\left[t\right]\right)|\left(0,1\right),\left(1,0\right),\left(1,1\right),\left(0,0\right)\}$. In each time slot, the decision is made by the mobile device, and the dynamic of the task buffer can be expressed as
\begin{equation}
q\left[t+1\right]=\min\{\left(q\left[t\right]-v_{L}\left[t\right]-v_{C}\left[t\right]\right)^{+}+a\left[t\right],Q\}, t=1,\cdots,
\label{Qdynamic}
\end{equation}
where $\left(x\right)^{+}\triangleq \max\{x,0\}$, $q\left[t\right]$ is the number of computation tasks in the buffer at the beginning of the $t$th time slot, and $a\left[t\right]\in\{0,1\}$ is the task arrival indicator, i.e., if a task arrives at the $t$th time slot, we have $a\left[t\right]=1$; otherwise, we have $a\left[t\right]=0$.

\subsection{Computation Model}
\textbf{i) Local Computation Model:} We assume that the CPU at the mobile device is operating at frequency $f_{loc}$ (in Hz) if a task is being executed, and its power consumption is given by $P_{loc}$ (in W); otherwise, the local CPU is idle and consumes no power. The number of required CPU cycles for executing a task successfully is denoted as $C$, which depends on the types of mobile applications \cite{Miettinen10}. In other words, $N=\lceil \frac{C}{f_{loc}\Delta}\rceil$ time slots are needed to complete a task. We use $c_{L}\left[t\right]\in \{0,1,\cdots,N-1\}$ to denote the processing state of the local CPU, where $c_{L}\left[t\right]=0$ means the local CPU is idle, while $c_{L}\left[t\right]= n$ ($1 \leq n \leq N-1$) indicates that one task is being processed at the local CPU, and $N-n$ more time slots are required to complete the task. For instance, $c_{L}\left[t\right]=N-1$ indicates that the task will be completed at the end of time slot $t$ and the local CPU will be available for a new task starting from the $\left(t+1\right)$th time slot.

\textbf{ii) Cloud Computation Model:} In order to offload a computation task to the MEC server, all the input data of the task should be successfully delivered to the MEC server over the wireless channel. Without loss of generality, we assume the input data of each task consists of $M$ equal-size data packets and each packet contains $R$ bits. For simplicity, on-off power control is adopted. We assume the channel side information is available at the mobile device, and thus a packet can be successfully transmitted to the MEC server if the achievable throughput in the $t$th time slot, $r\left(\gamma\left[t\right],P_{\rm{tx}}\right)$, is no smaller than the packet size, i.e.,
$r\left(\gamma\left[t\right],P_{\rm{tx}}\right)=B\log_{2}\left(1+\frac{\gamma\left[t\right] P_{\rm{tx}}}{N_{0}B}\right)\geq R$,
where $\gamma\left[t\right]$ is the channel power gain in the $t$th time slot, $P_{\rm{tx}}$ is the transmit power, $B$ is the system bandwidth and $N_{0}$ is the noise power spectral density at the receiver; otherwise, the transmitter will be silent and consume no power.

We use $c_{T}\left[t\right]\in\{0,1,\cdots, M\}$ to represent the  state of the TU, where  $c_{T}\left[t\right]=0$ means that the TU is available for offloading a task  to the MEC server, and $c_{T}\left[t\right]=m \left(1\leq m\leq M\right)$ means that the $m$th packet of one task is scheduled to transmit in time slot $t$.  When all the input bits are successfully received, the MEC server begins to execute the task. Assume that the MEC server is equipped with a multi-core CPU so that concurrent execution of multiple tasks is feasible. Similar to local computation, $N_{cloud}=\lceil \frac{C}{f_{ser}\Delta}\rceil$ time slots are required for completing the task at the MEC server, where $f_{ser}$ denotes the CPU-cycle frequency at the MEC server. Besides, we denote the delay for feeding back the computation results as $t_{r}$, which is viewed as a constant.

\section{Stochastic Computation Task Scheduling and Markov Chain Modeling}

In the considered system, the system state $\bm{\tau}\left[t\right]$ can be characterized by a triplet, i.e., $\bm{\tau}[t]=(q[t],c_{T}[t],c_{L}[t])$. Thus, the state space $\mathcal{S}$ can be expressed as $\mathcal{S}=\{0,1,\cdots,Q\}\times \{0,1,\cdots,M\}\times \{0,1,\cdots,N-1\}$, where ``$\times$'' denotes the Cartesian product. In the following, we will introduce the stochastic computation task scheduling policy, and analyze the average delay of each task and the average power consumption at the mobile device using Markov chain theory.

\subsection{Stochastic Computation Task Scheduling}
In order to minimize the average delay of each task and to meet the average power constraint, the mobile device should make the computation task scheduling decision at each time slot, i.e., whether to schedule a task for local computing or to offload it to the MEC server. To characterize the computation task scheduling policy, we introduce a set of probabilistic parameters $\{g_{\bm{\tau}}^{k}\}$ where $g_{\bm{\tau}}^{k}\in[0,1],\forall \bm{\tau}\in\mathcal{S},k=1,2,3,4$, which is a mapping from the system state to the probability space. The superscript $k$ in $\{g_{\bm{\tau}}^{k}\}$ indicates the four possible decisions as mentioned in Section II-A. In particular, $k=1,2,3,4$ refer to the computation task scheduling decision $(0,1),(1,0),(1,1),(0,0)$, respectively.

It is straightforward that each computation task can only be scheduled for local computing (cloud computing)  when the local CPU (TU) is idle. When the task buffer is empty, i.e., $q[t]=0$, there is no task to be scheduled, i.e., $g_{(0,m,n)}^{k}=0$ for $k=1,2,3$ and $g_{(0,m,n)}^{4}=1$. In the following, we consider the cases with $q\left[t\right]>0$ assuming different availabilities of the CPU and TU at the mobile device.

\textbf{Case I:} $c_{L}[t]=c_{T}[t]=0$. In this case, both the local CPU and the transmitter are idle. Thus, at most two computation tasks can start to be processed, i.e., one for local computing and the other for computation offloading. Given the system
state $\bm{\tau}\left[t\right]=(q[t],c_{T}[t],c_{L}[t])=(i,0, 0)$ ($i\geq2$), the computation task scheduling policy can be expressed as
\begin{equation}
(v_{C}[t],v_{L}[t])=\begin{cases}
(0,1) & w.p.\, g_{\bm{\tau}}^{1},\\
(1,0) & w.p.\, g_{\bm{\tau}}^{2},\\
(1,1) & w.p.\, g_{\bm{\tau}}^{3},\\
(0,0) & w.p.\,(1-\sum_{k=1}^{3}g_{\bm{\tau}}^{k}).
\end{cases}
\label{CASEI1}
\end{equation}
When there is only one task buffered in the queue, i.e., $i=1$, decision $\left(1,1\right)$ is infeasible and thus the computation task scheduling policy can be expressed as
\begin{equation}
(v_{C}[t],v_{L}[t])=\begin{cases}
(0,1) & w.p.\, g_{\bm{\tau}}^{1},\\
(1,0) & w.p.\, g_{\bm{\tau}}^{2},\\
(0,0) & w.p.\,(1-\sum_{k=1}^{2}g_{\bm{\tau}}^{k}).
\end{cases}
\label{CASEI2}
\end{equation}

\textbf{Case II:} $c_{L}[t]=0$ and $c_{T}[t]> 0$. The local CPU is idle while the transmitter is occupied. Thus, the mobile device can decide whether to start to compute one task locally or to remain idle. Hence, $\forall \bm{\tau}=(i,m,n)$ ($i\geq1$),
the computation task scheduling policy can be expressed as
\begin{equation}
(v_{C}[t],v_{L}[t])=\begin{cases}
(0,1) & w.p.\, g_{\bm{\tau}}^{1},\\
(0,0) & w.p.\,(1-g_{\bm{\tau}}^{1}).
\end{cases}
\label{CASEII}
\end{equation}

\textbf{Case III:} $c_{L}[t] >0$ and $c_{T}[t]=0$. The transmitter is idle and the mobile device  decides whether to send one computation task to the MEC server over the  wireless link. Hence, the task scheduling policy can be represented by
\begin{equation}
(v_{C}[t],v_{L}[t])=\begin{cases}
(1,0) & w.p.\, g_{\bm{\tau}}^{2},\\
(0,0) & w.p.\,(1-g_{\bm{\tau}}^{2}).
\end{cases}
\label{CASEIII}
\end{equation}

\textbf{Case IV:} $c_{L}[t]>0$ and {$c_{T}[t]>0$}. Both the local CPU and the TU are occupied, and  $\Pr\{(v_{C}[t],v_{L}[t])=(0,0)\}=g_{\bm{\tau}}^{4}=1$.

It is worthwhile to note that the performance of the MEC system depends on the adopted computation offloading policy, which can be characterized by the set of parameters $\{g_{\bm{\tau}}^{k}\}$ and the optimal computation offloading policy will be developed in Section IV.

\subsection{Delay and Power Analysis}
In this subsection, we will analyze the average delay of each task and the average power consumption at the mobile device by modeling the MEC system as a Markov chain.

Let $\chi_{\bm{\tau},\bm{\tau}{'}}=\Pr\{\bm{\tau}\rightarrow\bm{\tau}{'}\}$ denote the one-step state transition probability from state $\bm{\tau}$
to $\bm{\tau}{'}$.{\footnote{Please refer to Appendix A for the expressions and detailed analysis of the state transition probability.}}  It can be checked under a given computation task scheduling policy $\{g^{k}_{\bm{\tau}}\}$.
Thus, the steady-state distribution $\{\pi_{\bm{\tau}}\}$ can be obtained by solving the following linear equation set \cite{Ross14}:
\begin{equation}
\begin{cases}
&\sum_{\bm{\tau}{'}\in\mathcal{S}}\chi_{\bm{\tau}{'},\bm{\tau}}\pi_{\bm{\tau}{'}}=\pi_{\bm{\tau}},\forall \bm{\tau}\in\mathcal{S}\\
&\sum_{\bm{\tau}\in\mathcal{S}}\pi_{\bm{\tau}}=1.
\end{cases}
\label{eq:steady_state}
\end{equation}

\textbf{Average Delay:}
As each computation task experiences a waiting stage and a processing stage (either local or cloud computing) after its arrival, according to the Little's Theorem \cite{Ross14}, the average queueing delay can
be expressed as
\begin{equation}
t_{q}=\frac{1}{\alpha}\sum\limits _{i=0}^{Q}i\cdot\Pr\{q[t]=i\}=\frac{1}{\alpha}\sum\limits _{i=0}^{Q}i\sum\limits _{m=0}^{M}\sum\limits _{n=0}^{N-1}\pi_{(i,m,n)},\label{eq:queuing_time}
\end{equation}
where $\alpha$ denotes the task  arrival rate and $\Pr\{q[t]=i\}=\sum_{m=0}^{M+1}\sum_{n=0}^{N}\pi_{(i,m,n)}$. Recall that the local execution time for each task is $N$ time slots, and the processing time of cloud computing includes the time spent on transmitting the input data of the task $t_{tx}$, the execution time at the MEC server $N_{cloud}$, as well as the time of feeding back the computation result $t_{rx}$, i.e.,
\begin{equation}
t_{c}=t_{tx}+N_{cloud}+t_{rx},\label{eq:time_cloud}
\end{equation}
where the average transmission time for each computation task is given by
\begin{equation}
t_{tx}=M\sum_{j=1}^{\infty}j(1-\beta)^{(j-1)}\beta. \label{eq:tx_time}
\end{equation}
In (\ref{eq:tx_time}), $\beta \triangleq {\rm{Pr}}\{r\left(\gamma\left[t\right],P_{\rm{tx}}\right)\geq R\}$ denotes the probability that the channel in not in outage. Consequently, the average processing time of each task can be expressed as
\begin{equation}
t_{p}=\eta N+(1-\eta)t_{c},\label{eq:processing_time}
\end{equation}
where $\eta$ denotes the proportion of computation tasks that are executed locally at the mobile device in the long run and can be computed according to the following equation:
\begin{equation}
\eta=\frac{\sum_{\bm{\tau}\in\mathcal{S}_{1}}\pi_{\bm{\tau}}g_{\bm{\tau}}^{1}+\sum_{\bm{\tau}\in\mathcal{S}_{3}}\pi_{\bm{\tau}}g_{\bm{\tau}}^{3}}{\sum_{\bm{\tau}\in\mathcal{S}_{1}}\pi_{\bm{\tau}}g_{\bm{\tau}}^{1}+\sum_{\bm{\tau}\in\mathcal{S}_{2}}\pi_{\bm{\tau}}g_{\bm{\tau}}^{2}+2\sum_{\bm{\tau}\in\mathcal{S}_{3}}\pi_{\bm{\tau}}g_{\bm{\tau}}^{3}},\label{eq:prob_eta}
\end{equation}
where the state sets $\mathcal{S}_{k}(k=1,2,3)$ are defined as $\mathcal{S}_{1}=\{(i,m,0)|i\geq1,m\in\{0,\cdots,M\}\}
 $, $\mathcal{S}_{2}=\{(i,0,n)|i\geq1,n\in\{0,\cdots,N-1\}\}$
  and $\mathcal{S}_{3}=\{(i,0,0)|i\geq2\}$, respectively.
Therefore, the average delay of each computation task is the sum of the queueing delay and the processing latency, which can be written as
\begin{equation}
\bar{T}=t_{q}+t_{p}.
\end{equation}

\textbf{Average Power Consumption:}

Let $\mu_{\bm{\tau}}^{loc}$ and $\mu_{\bm{\tau}}^{tx}$ denote the probabilities of local computations and successful packet transmissions with power consumptions $P_{loc}$ and $P_{tx}$, respectively, given the system state $\bm{\tau}=(i,m,n)$. Thus, the average power consumption at the mobile device
is given by
\begin{equation}
\bar{P}=\sum\limits _{\bm{\tau}\in\mathcal{S}}\pi_{\bm{\tau}}\left(\mu_{\bm{\tau}}^{loc}P_{loc}+\mu_{\bm{\tau}}^{tx}P_{tx}\right),\label{eq:power_consumption}
\end{equation}
where the power coefficients $\mu_{\bm{\tau}}^{loc}$ and $\mu_{\bm{\tau}}^{tx}$ for each state $\bm{\tau}$ can be expressed as
\begin{equation}
\mu_{\bm{\tau}}^{loc}=\begin{cases}
g_{\bm{\tau}}^{1}+g_{\bm{\tau}}^{3}, & \bm{\tau}=(i,0,0)\,(\forall i\geq2)\\
g_{\bm{\tau}}^{1}, & \bm{\tau}=(1,0,0)\cup(i,m,0)(\forall i\geq1,m>0)\\
1, & \bm{\tau}=(i,m,n)\,(\forall i\geq0,m\geq0,n>0)\\
0, & \text{otherwise}
\end{cases}\label{eq:mu_loc}
\end{equation}
and
\begin{equation}
\mu_{\bm{\tau}}^{tx}=\begin{cases}
\beta(g_{\bm{\tau}}^{2}+g_{\bm{\tau}}^{3}), & \bm{\tau}=(i,0,0)\,(\forall i\geq2)\\
\beta g_{\bm{\tau}}^{2}, & \bm{\tau}=(1,0,0)\cup(i,0,n)\,(\forall i\geq1,n>0)\\
\beta, & \bm{\tau}=(i,m,n)\,(\forall i\geq0,m>0,n\geq0)\\
0, & \text{otherwise},
\end{cases}\label{eq:mu_tx}
\end{equation}
respectively. The derivation of the  power coefficients $\mu_{\bm{\tau}}^{loc}$ and $\mu_{\bm{\tau}}^{tx}$ is deferred to Appendix \ref{sec:power_cof_state}. Therefore, by averaging over all the state $\{\bm{\tau}=(i,m,n)\in\mathcal{S}\}$, we have   $\bar{P}=\nu_{loc}P_{loc}+\nu_{tx}P_{tx}$ with the average power coefficients given by  $\nu_{loc}=\sum\limits _{\bm{\tau}\in\mathcal{S}}\pi_{\bm{\tau}}\mu_{\bm{\tau}}^{loc}$
and $\nu_{tx}=\sum\limits _{\bm{\tau}\in\mathcal{S}}\pi_{\bm{\tau}}\mu_{\bm{\tau}}^{tx}$, respectively.

\section{Optimal Computation Offloading Scheduling}

In this section, we will formulate an optimization problem to minimize the average delay of each computation task subject to the average power constraint at the mobile device. An optimal algorithm will then be developed for the formulated optimization problem.

Based on the delay and power analysis in Section III-B, the power-constrained delay minimization problem can be formulated in $\mathcal{P}_{1}$:
\begin{equation}
\begin{array}[t]{cl}
\mathcal{P}_{1}:\min\limits_{\{g_{\bm{\tau}}^{k}\}} & \bar{T}=\frac{1}{\alpha}\sum\limits _{i=0}^{Q}i\sum\limits _{m=0}^{M}\sum\limits _{n=0}^{N-1}\pi_{(i,m,n)}+\eta N+(1-\eta)t_{c}\\
\ \ \ \ {\rm{s.t.}} & \begin{cases}
\bar{P}\leq\bar{P}_{max}, &\ \ \ \ \ \ \ \ \ \ \ (\mathrm{a})\\
\sum_{\bm{\tau}{'}\in\mathcal{S}}\chi_{\bm{\tau}{'},\bm{\tau}}\pi_{\bm{\tau}{'}}=\pi_{\bm{\tau}},\,\bm{\tau}\in\mathcal{S}, &\ \ \ \ \ \ \ \ \ \ \  (\mathrm{b})\\
\sum\limits _{i=0}^{Q}\sum\limits _{m=0}^{M}\sum\limits _{n=0}^{N-1}\pi_{(i,n,m)}=1, &\ \ \ \ \ \ \ \ \ \ \  (\mathrm{c})\\
\sum_{k=1}^{4}g_{(i,m,n)}^{k}=1,\,\forall i,m,n, &\ \ \ \ \ \ \ \ \ \ \ (\mathrm{d})\\
g_{(i,m,n)}^{k}\geq0,\,\forall i,m,n,k, &\ \ \ \ \ \ \ \ \ \ \ (\mathrm{e})
\end{cases}
\end{array}\label{eq:Opt_problem}
\end{equation}
where (\ref{eq:Opt_problem}.a) is the average power constraint, (\ref{eq:Opt_problem}.b) and  (\ref{eq:Opt_problem}.c) denote the balance equation set, and $\eta$ is given by (\ref{eq:prob_eta}). It is worthwhile to note that once $\{g_{\bm{\tau}}^{k}\}$ is determined, $\pi_{\bm{\tau}}$ can be obtained according to (\ref{eq:steady_state}). However, as $\mathcal{P}_{1}$ is non-convex, the optimal solution is not readily available. In the following, we will reformulate $\mathcal{P}_{1}$ into a series of linear programming (LP) problems in order to obtain its optimal solution. First, we define the occupation measure $\{x_{\bm{\tau}}^{k}\}$  as $x_{\bm{\tau}}^{k}=\pi_{\bm{\tau}}g_{\bm{\tau}}^{k}$, which is the probability that the system is in state $\bm{\tau}=\left(i,m,n\right)$ while decision $k$ is made  \cite{EAltman99}. By definition, $\sum_{k=1}^{4}g_{\bm{\tau}}^{k}=1$, and thus $\pi_{\bm{\tau}}=\sum_{k=1}^{4}x_{\bm{\tau}}^{k}$.

By replacing $\{\pi_{\left(i,m,n\right)}\}$ with $\{x_{(i,m,n)}^{k}\}$ in $\mathcal{P}_{1}$, we obtain an equivalent formulation of $\mathcal{P}_{1}$ as follows:
\begin{equation}
\begin{array}[t]{cl}
\mathcal{P}_{2}:\min\limits _{\bm{x},\eta} & \bar{T}=\frac{1}{\alpha}\sum\limits_{\bm{\tau}\in\mathcal{S}}\sum\limits _{k=1}^{4}i\cdot x_{\bm{\tau}}^{k}+\eta N+(1-\eta)t_{c}\\
\ \ \ \ {\rm{s.t.}} & \begin{cases}
\nu_{loc}(\bm{x})P_{loc}+\beta\nu_{tx}(\bm{x})P_{tx}\leq\bar{P}_{max}, &\ \ \ \ \  (\mathrm{a})\\
\Gamma(\bm{x},\eta)=0, &\ \ \ \ \ (\mathrm{b})\\
F_{\bm{\tau}}(\bm{x})=0,\,\forall\bm{\tau}=(i,m,n)\in\mathcal{S}, &\ \ \ \ \  (\mathrm{c})\\
\sum\limits _{i=0}^{Q}\sum\limits _{m=0}^{M}\sum\limits _{n=0}^{N-1}\sum\limits _{k=1}^{4}x_{(i,m,n)}^{k}=1, &\ \ \ \ \  (\mathrm{d})\\
x_{(i,m,n)}^{k}\geq0,\forall i,m,n,k,\,\eta\in[0,1], &\ \ \ \ \ (\mathrm{e})
\end{cases}
\end{array}\label{eq:LP_problem}
\end{equation}
where $\nu_{loc}(\bm{x})$ and $\nu_{tx}(\bm{x})$ are linear functions of the variables $\bm{x}$ given by
\begin{equation}
\begin{split}\nu_{loc}(\bm{x})= & \sum_{i=1}^{Q}x_{(i,0,0)}^{1}+\sum_{i=2}^{Q}x_{(i,0,0)}^{3}+\sum_{i\geq1}\sum_{m=1}^{M}x_{(i,m,0)}^{1}\\
 & +\sum_{i\geq0}\sum_{m=0}^{M}\sum_{n=1}^{N-1}\sum_{k=1}^{4}x_{(i,m,n)}^{k},
\end{split}
\end{equation}
and
\begin{equation}
\begin{split}\nu_{loc}(\bm{x})= & \sum_{i=1}^{Q}x_{(i,0,0)}^{1}+\sum_{i=2}^{Q}x_{(i,0,0)}^{3}+\sum_{i\geq1}\sum_{n=1}^{N-1}x_{(i,0,n)}^{2}\\
 & +\sum_{i\geq0}\sum_{m=1}^{M}\sum_{n=0}^{N-1}\sum_{k=1}^{4}x_{(i,m,n)}^{k},
\end{split}
\end{equation}
respectively, and
$\Gamma(\bm{x},\eta)$ and $F_{\bm{\tau}}(\bm{x})$ can  be expressed as
\begin{equation}
\Gamma(\bm{x},\eta)=  (1-\eta)\sum\limits _{\bm{\tau}\in\mathcal{S}_{1}}x_{\bm{\tau}}^{1}-\eta\sum\limits _{\bm{\tau}\in\mathcal{S}_{2}}x_{\bm{\tau}}^{2}
 +(1-2\eta)\sum\limits _{\bm{\tau}\in\mathcal{S}_{3}}x_{\bm{\tau}}^{3},
\end{equation}
and
\begin{equation}
F_{\bm{\tau}}(\bm{x})=\sum_{\bm{\tau}^{'}\in\mathcal{S}}\sum_{k=1}^{4}\tilde{\chi}_{\bm{\tau}^{'},\bm{\tau},k}x_{\bm{\tau}^{'}}^{k}-\sum_{k=1}^{4}x_{\bm{\tau}}^{k},
\end{equation}
respectively.\footnote{$\tilde{\chi}_{\bm{\tau}',\bm{\tau},k}$ denotes the probability that the current system state is $\bm{\tau}'$ and decision $k$ is made, while the system state in the next time slot is $\bm{\tau}$, which is independent with $\{g_{\bm{\tau}}^{k}\}$ in contrast to $\chi_{\bm{\tau}',\bm{\tau}}$. Note that $\sum_{k=1}^{4}\tilde{\chi}_{\bm{\tau}',\bm{\tau},k}g_{\bm{\tau}'}^{k}=\chi_{\bm{\tau}',\bm{\tau}},\forall \bm{\tau}',\bm{\tau}$.}

The optimal solution and the optimal value of $\mathcal{P}_{2}$ are  denoted as $(\bm{x}^{*},\eta^{*})$ and $\bar{T}'\left(\eta^*\right)$, respectively. Once $\bm{x}^{*}$ is known, the optimal computation task scheduling policy $\{g_{\bm{\tau}}^{k*}\}$ can be obtained as
\begin{equation}
g_{\bm{\tau}}^{k*}=\frac{x_{\bm{\tau}}^{k*}}{{\sum\nolimits_{k=1}^{4}x_{\bm{\tau}}^{k*}}},\forall \bm{\tau}\in\mathcal{S},k\in\{1,2,3,4\}.
\end{equation}

Due to the product form of $\eta$ and $x_{\left(i,m,n\right)}^{k}$ in (\ref{eq:LP_problem}.b), $\mathcal{P}_{2}$ is still a non-convex problem.
Fortunately, we observe that for a given value of $\eta$, $\mathcal{P}_{2}$ reduces to an LP problem in terms of variables $\{x_{\bm{\tau}}^{k}\}$. Therefore, we can first obtain the optimal solution $\bm{x}{'}(\eta)$ for arbitrary $\eta\in\left[0,1\right]$ and conduct a one-dimensional search for the optimal $\eta^{*}$. Detailed procedures for solving $\mathcal{P}_{2}$ are summarized in Algorithm \ref{alg:LP_linesearch}.

\begin{algorithm}[t]
\protect\caption{A one-dimensional search algorithm for solving $\mathcal{P}_{2}$ \label{alg:LP_linesearch}}

\begin{algorithmic}[1]

\STATE Set $\eta=0$ and $J$ as a sufficiently large integer;

\FOR{$j=0:1:J$}

\STATE Solve the LP problem (\ref{eq:LP_problem}) with a fixed $\eta$;

\STATE Obtain the optimal solution $\bm{x}{'}(\eta)$ and  the optimal value  $\bar{T}{'}(\eta)$;

\STATE Update the variable $\eta=\eta+1/J$;

\ENDFOR

\STATE Find the optimal solution ($\bm{x}^*,\eta^*$) with $\eta^{*}=\arg\min_{\eta}\bar{T}{'}(\eta)$
and $\bm{x}^{*}=\bm{x}{'}(\eta^{*})$.

\end{algorithmic}
\end{algorithm}

\section{Simulation Results }

In this section, we evaluate the performance of the proposed stochastic computation task scheduling  policy by simulations. In simulations, we assume that the input data  size of each task is $500$ Kbits and $C=1300\times L$ CPU cycles \cite{Miettinen10}. The  path-loss constant is set to be   $1.6\times10^{-7}$. Each task is encapsulated into one packet, and thus $M=1$ and $R=\frac{L}{M\Delta}$ bits. In addition, we set $\Delta=20$ ms, $B=5$ MHz, $P_{tx}=1$ W, $\sigma^{2}=N_{0}B=10^{-9}$ W, $f_{loc}=2$ GHz, and $f_{ser}=100$ GHz. The time $t_{rx}$ is approximated as zero, and thus   $\beta=0.4$, $t_{tx}=2.5$, $N=17$, $P_{l}\approx10^{-28}\times f_{loc}^{3}=0.8$ W, $N_{cloud}=1$, and $t_c=t_{tx}+N_{cloud}+t_{rx}=3.5$.

\begin{figure}[h]
\centering
\renewcommand{\figurename}{Fig.}\includegraphics[width=0.45\textwidth]{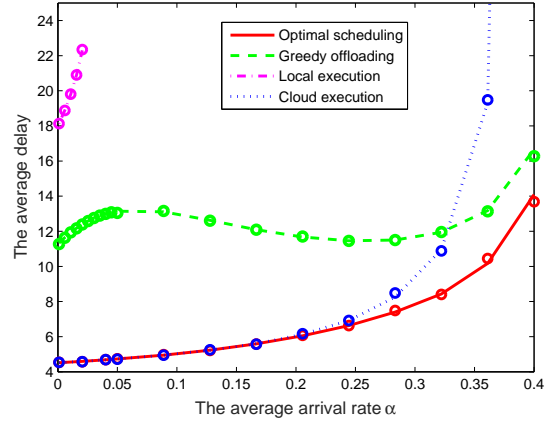} \protect
\caption{The average delay vs. the average arrival rate. \label{fig:delay_arrival}}
\end{figure}

\begin{figure}[h]
\centering
\renewcommand{\figurename}{Fig.}\includegraphics[width=0.45\textwidth]{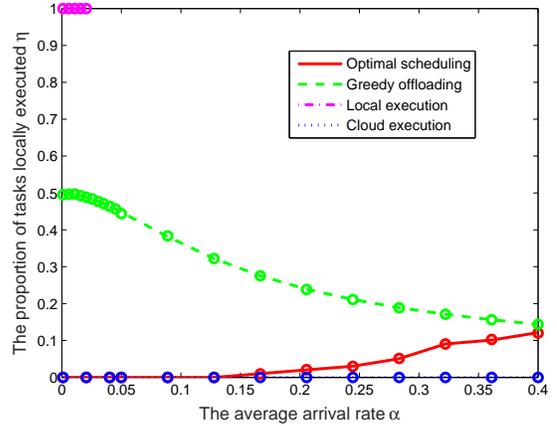} \protect
\caption{The proportion of tasks that are executed locally vs. the average arrival rate. \label{fig:eta_arrival}}
\end{figure}

We introduce three baseline task scheduling  policies, including the \textbf{local execution policy}, which executes all the computation tasks locally at the mobile device; the \textbf{cloud execution policy}, where all the tasks are offloaded to the MEC server for cloud computing; and the \textbf{greedy offloading policy}, where the mobile device schedules  the computation tasks waiting in the buffer  to the local CPU   or the MEC server for task executions   whenever the local CPU or TU is idle.


The average delay $\bar{T}$ and the proportion of computation tasks that are executed locally at the mobile device $\eta$ achieved by different computation task scheduling policies are shown in Fig.$\,$\ref{fig:delay_arrival} and Fig.$\,$\ref{fig:eta_arrival}, respectively. It can be observed from the figures that, the average delays (proportions of tasks executed locally) achieved by the local execution, cloud execution and the optimal policy, increase with the average computation task arrival rate $\alpha$, which is in accordance with our intuition. In contrast, the behavior of the greedy offloading policy is sharply different from that of the other three policies: less computation tasks are scheduled to the local CPU, and $\eta$  decreases with $\alpha$, as shown in Fig.$\,$\ref{fig:eta_arrival}. This is due to the fact that the execution time required by MEC is much smaller than that by the local CPU, i.e., $t_c=3.5 <N=17$. When $\alpha$ increases, more arrived tasks should be sent to the MEC server for faster  completion to adapt to the increasing task arrival rate. Consequently, the average delay achieved by the greedy offloading policy fluctuates since the average queueing delay increases with $\alpha$, while the processing time of each task $t_p$ decreases with $\alpha$.

When the local (cloud) execution policy is applied, $\eta$ equals  $1$ ($0$), and hence the execution time $t_p$ (c.f. \eqref{eq:processing_time}) is a constant. Thus,  the average delay $\bar {T}=t_q+t_p$ is solely determined by  the average queueing delay and increases with the task arrival rate. When the optimal offloading policy is applied, $\eta =0$ when $\alpha\leq 0.12$ as shown in Fig.$\,$\ref{fig:eta_arrival}. In this case, the mobile device prefers to schedule all the tasks for cloud computing since the cloud execution time $t_c$ is much smaller than the local execution time  $N$ and no parallel execution is needed. With $\alpha$ grows beyond 0.12, more tasks should be sent to the local CPU  in a way that parallel local and cloud executions can be fully exploited to adapt to the increased task arrival rate.  Therefore, the average delay increases with the task arrival rate due to the increase of the processing time and queueing delay. When $\alpha$ approaches $0.4$, the behaviour of the optimal scheduling policy converges to the greedy offloading policy. Among all the policies, the proposed optimal scheduling policy achieves the minimum
average delay, and meanwhile, it enjoys the largest stable region, i.e., it is capable of accommodating the maximum task arrival rate $\alpha$. This is because  our proposed task scheduling policy employs parallel local and cloud executions in a delay-optimal manner.

\section{Conclusions}

In this paper, we proposed a stochastic computation task scheduling policy for MEC systems, incorporating different timescales in the task execution process and the channel fading  process. Based on the analysis of the average delay and the average power consumption at the mobile device, we developed an efficient one-dimensional search algorithm to find the optimal task scheduling policy. It was found that our proposed stochastic task scheduling policy achieves the minimum average delay in various scenarios compared to three baseline policies. For future investigation, it would be interesting to extend this work to more general MEC systems.

\appendix
\subsection{The State Transition Probabilities of the Markov Chain \label{sec:transition_prob}}
In this subsection, we will discuss the state transition probabilities of the Markov chain in the following four cases:

\textbf{ Case I:} $c_{T}[t]=c_{L}[t]=0$.
In this case, both the local CPU and the TU  are idle and each of them is available for processing a new task. When at least two computation tasks are waiting in the task buffer, one of four computation task scheduling decisions can be chosen with probability $g_{\bm{\tau}}^{k}, k=1,\cdots,4$, as presented in (\ref{CASEI1}).  By jointly considering all possible task arrival states and channel states, for any given $\bm{\tau}=(i,0,0)$ ($\forall i\geq2$), the state transition probabilities can be expressed as
\begin{equation}
\begin{cases}
\Pr\{(i,0,\Gamma_{N}(1))|(i,0,0)\}=\alpha g_{\bm{\tau}}^{1},\\
\Pr\{(i,1,0)|(i,0,0)\}=\alpha(1-\beta)g_{\bm{\tau}}^{2},\\
\Pr\{(i,\Gamma_{M+1}(2),0)|(i,0,0)\}=\alpha\beta g_{\bm{\tau}}^{2},\\
\Pr\{(i-1,1,\Gamma_{N}(1))|(i,0,0)\}=\alpha(1-\beta)g_{\bm{\tau}}^{3},\\
\Pr\{(i-1,\Gamma_{M+1}(2),\Gamma_{N}(1))|(i,0,0)\}=\alpha\beta g_{\bm{\tau}}^{3},\\
\Pr\{(i+1,0,0)|(i,0,0)\}=\alpha(1-\sum\limits _{k=1}^{3}g_{\bm{\tau}}^{k}),\\
\Pr\{(i-1,0,\Gamma_{N}(1))|(i,0,0)\}=(1-\alpha)g_{\bm{\tau}}^{1},\\
\Pr\{(i-1,1,0)|(i,0,0)\}=(1-\alpha)(1-\beta)g_{\bm{\tau}}^{2},\\
\Pr\{(i-1,\Gamma_{M+1}(2),0)|(i,0,0)\}=(1-\alpha)\beta g_{\bm{\tau}}^{2},\\
\Pr\{(i-2,1,\Gamma_{N}(1))|(i,0,0)\}=(1-\alpha)(1-\beta)g_{\bm{\tau}}^{3},\\
\Pr\{(i-2,\Gamma_{M+1}(2),\Gamma_{N}(1))|(i,0,0)\}=(1-\alpha)\beta g_{\bm{\tau}}^{3},\\
\Pr\{(i,0,0)|(i,0,0)\}=(1-\alpha)(1-\sum\limits _{k=1}^{3}g_{\bm{\tau}}^{k}),
\end{cases}
\end{equation}
where the state mapping function $\Gamma_{M}(m)$ is defined as
\begin{equation}
\Gamma_{M}(m)=\begin{cases}
m, & m\in\{0,1,\cdots,M-1\},\\
0, & m=M.
\end{cases}
\end{equation}
For example, state $(i,0,0)$ will transfer to state $(i,0, \Gamma_{N}\left(1\right))$ with probability $\alpha g_{\bm{\tau}}^1$, when one new task arrives at the task buffer and one waiting task is sent to the local CPU.

When there is just one task in the task buffer, the mobile device has three possible decisions: local execution, cloud execution and remaining idle, as given by (\ref{CASEI2}). Accordingly, for $\bm{\tau}=(1,0,0)$, the state transition probability can be written as

\begin{equation}
\begin{cases}
\Pr\{(i,0,\Gamma_{N}(1))|(1,0,0)\}=\alpha g_{\bm{\tau}}^{1},\\
\Pr\{(i,1,0)|(1,0,0)\}=\alpha(1-\beta)g_{\bm{\tau}}^{2},\\
\Pr\{(i,\Gamma_{M+1}(2),0)|(1,0,0)\}=\alpha\beta g_{\bm{\tau}}^{2},\\
\Pr\{(i+1,0,0)|(1,0,0)\}=\alpha(1-\sum\limits _{k=1}^{2}g_{\bm{\tau}}^{k}),\\
\Pr\{(i-1,0,\Gamma_{N}(1))|(1,0,0)\}=(1-\alpha)g_{\bm{\tau}}^{1},\\
\Pr\{(i-1,1,0)|(1,0,0)\}=(1-\alpha)(1-\beta)g_{\bm{\tau}}^{2},\\
\Pr\{(i-1,\Gamma_{M+1}(2),0)|(1,0,0)\}=(1-\alpha)\beta g_{\bm{\tau}}^{2},\\
\Pr\{(i,0,0)|(1,0,0)\}=(1-\alpha)(1-\sum\limits _{k=1}^{2}g_{\bm{\tau}}^{k}),
\end{cases}
\end{equation}
by jointly considering different computation task scheduling decisions, task arrival and channel states.

When the task buffer is empty, neither local execution nor cloud execution is needed. In this case, the system state transits due to one new task arrival, and accordingly the state transition probability can be simplified as
\begin{equation}
\begin{cases}
\Pr\{(1,0,0)|(0,0,0)\} & =\alpha,\\
\Pr\{(0,0,0)|(0,0,0)\} & =(1-\alpha).
\end{cases}
\end{equation}

\textbf{ Case II:}  $c_{T}[t]>0$ and $c_{L}[t]=0$.
In this case, the local CPU is available to execute a new task while the task offloading is in process. When there is at least one packet in the task buffer, i.e.,  $\bm{\tau}=(i,m,0)$ ($\forall i\geq1,m\in\{1,\cdots,M\}$), the computation task scheduling policy is given by (\ref{CASEII}). Accordingly, the
state transition probabilities can be written as

\begin{equation}
\begin{cases}
\Pr\{(i,\Gamma_{M+1}(m+1),\Gamma_{N}(1))|(i,m,0)\}=\alpha\beta g_{\bm{\tau}}^{1},\\
\Pr\{(i,m,\Gamma_{N}(1))|(i,m,0)\}=\alpha(1-\beta)g_{\bm{\tau}}^{1},\\
\Pr\{(i+1,\Gamma_{M+1}(m+1),0)|(i,m,0)\}=\alpha\beta(1-g_{\bm{\tau}}^{1}),\\
\Pr\{(i+1,m,0)|(i,m,0)\}=\alpha(1-\beta)(1-g_{\bm{\tau}}^{1}),\\
\Pr\{(i-1,\Gamma_{M+1}(m+1),\Gamma_{N}(1))|(i,m,0)\}=(1-\alpha)\beta g_{\bm{\tau}}^{1},\\
\Pr\{(i-1,m,\Gamma_{N}(1))|(i,m,0)\}=(1-\alpha)(1-\beta)g_{\bm{\tau}}^{1},\\
\Pr\{(i,\Gamma_{M+1}(m+1),0)|(i,m,0)\}=(1-\alpha)\beta(1-g_{\bm{\tau}}^{1}),\\
\Pr\{(i,m,0)|(i,m,0)\}=(1-\alpha)(1-\beta)(1-g_{\bm{\tau}}^{1}).
\end{cases}
\end{equation}

When the task buffer is empty, there exist four possible state transitions with their transition probabilities given by
\begin{equation}
\begin{cases}
\Pr\{(1,\Gamma_{M+1}(m+1),0)|(0,m,0)\}=\alpha\beta,\\
\Pr\{(1,m,0)|(0,m,0)\}=\alpha(1-\beta),\\
\Pr\{(0,\Gamma_{M+1}(m+1),0)|(0,m,0)\}=(1-\alpha)\beta,\\
\Pr\{(0,m,0)|(0,m,0)\}=(1-\alpha)(1-\beta),
\end{cases}
\end{equation}
depending on whether there are one new task arrival and one successful packet delivery.

\textbf{ Case III:}  $c_{T}[t]=0$ and $c_{L}[t]>0$.
When the local CPU is busy in task execution  while the TU is idle, the decision on task offloading is made
with probability $g_{\bm{\tau}}^{2}$ when the task buffer is non-empty, as shown in (\ref{CASEIII}).
Similarly, the state transition probabilities can be obtained as

\begin{equation}
\begin{cases}
\Pr\{(i,1,\Gamma_{N}(n+1))|(i,0,n)\}=\alpha(1-\beta)g_{\bm{\tau}}^{2},\\
\Pr\{(i,\Gamma_{M+1}(2),\Gamma_{N}(n+1))|(i,0,n)\}=\alpha\beta g_{\bm{\tau}}^{2},\\
\Pr\{(i+1,0,\Gamma_{N}(n+1))|(i,0,n)\}=\alpha(1-g_{\bm{\tau}}^{2}),\\
\Pr\{(i-1,1,\Gamma_{N}(n+1))|(i,0,n)\}=(1-\alpha)(1-\beta)g_{\bm{\tau}}^{2},\\
\Pr\{(i-1,\Gamma_{M+1}(2),\Gamma_{N}(n+1))|(i,0,n)\}=(1-\alpha)\beta g_{\bm{\tau}}^{2},\\
\Pr\{(i,0,\Gamma_{N}(n+1))|(i,0,n)\}=(1-\alpha)g_{\bm{\tau}}^{2},
\end{cases}
\end{equation}
for $\bm{\tau}=(i,0,n)$ ($\forall i>0,n\in\{1,\cdots,N-1\}$).

When the task queue is empty, the state transition probabilities can be written as

\begin{equation}
\begin{cases}
\Pr\{(1,0,\Gamma_{N}(n+1))|(0,0,n)\} & =\alpha,\\
\Pr\{(0,0,\Gamma_{N}(n+1))|(0,0,n)\} & =(1-\alpha),
\end{cases}
\end{equation}
for $\bm{\tau}=(0,0,n)$ ($\forall n\in\{1,\cdots,N-1\}$). In this case, the system state transits due to the new task arrival and the naturally evolving local computing state.

\textbf{ Case IV:} $c_{T}[t]>0$ and $c_{L}[t]>0$.
In this case, both of the local CPU and the TU
are busy in processing. For $\bm{\tau}=(i,m,n)$ ($\forall i\geq0$, $m\in\{1,\cdots,M\}$,$n\in\{1,\cdots,N-1\}$), there also exist four possible state transitions with their transition probabilities given by

\begin{equation}
\begin{cases}
\Pr\{(i+1,\Gamma_{M+1}(m+1),\Gamma_{N}(n+1))|(i,m,n)\}=\alpha\beta,\\
\Pr\{(i+1,m,\Gamma_{N}(n+1))|(i,m,n)\}=\alpha(1-\beta),\\
\Pr\{(i,\Gamma_{M+1}(m+1),\Gamma_{N}(n+1))|(i,m,n)\}=(1-\alpha)\beta,\\
\Pr\{(i,m,\Gamma_{N}(n+1))|(i,m,n)\}=(1-\alpha)(1-\beta).
\end{cases}
\end{equation}

Notice that in some special cases, the destination states are exactly the same, and therefore the probabilities of transition to the common destination states should be combined.

\subsection{The Power Coefficients $\mu_{\bm{\tau}}^{loc}$ and $\mu_{\bm{\tau}}^{tx}$   \label{sec:power_cof_state} }
In this subsection, we will derive the power coefficients
$\mu_{\bm{\tau}}^{loc}$ and $\mu_{\bm{\tau}}^{tx}$ based on the stochastic computation task scheduling policy described in Section III.

\textbf{  Case I:} $c_{T}[t]=c_{L}[t]=0$.
When there are at least two computation tasks in the task buffer, local execution and task offloading are conducted with probabilities
$(g_{\bm{\tau}}^{1}+g_{\bm{\tau}}^{3})$ and $\beta(g_{\bm{\tau}}^{2}+g_{\bm{\tau}}^{3})$,
respectively. Thus, for $\bm{\tau}=(i,0,0)$ ($\forall i\geq 2$), the power coefficients, i.e., the probabilities of consuming the execution power $P_{loc}$ and transmission power $P_{tx}$, are given by $\mu_{\bm{\tau}}^{loc}=g_{\bm{\tau}}^{1}+g_{\bm{\tau}}^{3}$ and $\mu_{\bm{\tau}}^{tx}=\beta(g_{\bm{\tau}}^{2}+g_{\bm{\tau}}^{3})$,
respectively. When the system is in state $\bm{\tau}=(1,0,0)$, the local execution (task offloading) is conducted with probability $g_{\bm{\tau}}^{1}$
$(\beta g_{\bm{\tau}}^{2})$. Thus, we have $\mu_{\bm{\tau}}^{loc}=g_{\bm{\tau}}^{1}$ and $\mu_{\bm{\tau}}^{tx}=\beta g_{\bm{\tau}}^{2}$.
When the task buffer is empty, i.e., $\bm{\tau}=(0,0,0)$, the power coefficients are $\mu_{\bm{\tau}}^{loc}=\mu_{\bm{\tau}}^{tx}=0$ since no power is consumed for neither local execution nor task offloading.

\textbf{Case II:} $c_{T}[t]>0$ and $c_{L}[t]=0$.
When the task buffer is non-empty, a computation task can be scheduled for local execution with probability $g_{\bm{\tau}}^{1}$  and
a packet of one task will be successfully transmitted to the MEC server with probability $\beta$. Therefore, the power coefficients are equal
to $\mu_{\bm{\tau}}^{loc}=g_{\bm{\tau}}^{1}$ and $\mu_{\bm{\tau}}^{tx}=\beta$. When the task buffer is empty, the power coefficients are equal to
$\mu_{\bm{\tau}}^{loc}=0$ and $\mu_{\bm{\tau}}^{tx}=\beta$ for $\bm{\tau}=(0,m,0)$ ($\forall m\in\{1,\cdots,M\}$).

\textbf{Case III:} $c_{T}[t]=0$ and $c_{L}[t]>0$. In this case, the local CPU is executing one task while the TU is available for delivering one packet of a new task successfully delivered with probability $\beta g_{\bm{\tau}}^{2}$ when the task buffer is non-empty.
Hence, we have $\mu_{\bm{\tau}}^{loc}=1$ and $\mu_{\bm{\tau}}^{tx}=\beta g_{\bm{\tau}}^{2}$
for $\bm{\tau}=(i,0,n)$ ($\forall i>0, n\in\{1,\cdots,N-1\}$). When the task buffer is empty, the power coefficients are equal to  $\mu_{\bm{\tau}}^{loc}=1$ and $\mu_{\bm{\tau}}^{tx}=0$ for $\bm{\tau}=(0,0,n)$ ($\forall n\in\{1,\cdots,N-1\}$).

\textbf{ Case IV:} $c_{T}[t]>0$ and $c_{L}[t]>0$. The power coefficients are $\mu_{\bm{\tau}}^{loc}=1$ and $\mu_{\bm{\tau}}^{tx}=\beta$,
since the local CPU is busy and one packet of a task will be successfully delivered with probability $\beta$.

Based on the above discussions, we can summarize the power coefficients $\mu_{\bm{\tau}}^{loc}$ and $\mu_{\bm{\tau}}^{tx}$ for each state $\bm{\tau}$ in \eqref{eq:mu_loc} and \eqref{eq:mu_tx}, respectively.

\end{document}